\documentclass[11pt]{article}
\usepackage{latexsym}  
\usepackage{amssymb}
\usepackage{graphicx}
\usepackage{amsmath}

\begin{document}

\hspace*{11cm} {MISC-2012-19}

\begin{center}
{\Large\bf Neutrino Mass Matrix Model with a Bilinear Form}

\vspace{4mm}
{\bf Yoshio Koide$^a$ and Hiroyuki Nishiura$^b$}

${}^a$ {\it Department of Physics, Osaka University, 
Toyonaka, Osaka 560-0043, Japan} \\
{\it E-mail address: koide@het.phys.sci.osaka-u.ac.jp}

${}^b$ {\it Faculty of Information Science and Technology, 
Osaka Institute of Technology, 
Hirakata, Osaka 573-0196, Japan}\\
{\it E-mail address: nishiura@is.oit.ac.jp}

\date{\today}
\end{center}

\vspace{3mm}

\begin{abstract} 
A neutrino mass matrix model with a bilinear form
$M_\nu = k_\nu (M_D M_R^{-1} M_D^T)^2$ is proposed 
within the framework of the so-called yukawaon model, which has 
been  proposed for the purpose of a unified description of 
the lepton mixing matrix $U_{PMNS}$ 
and the quark mixing matrix $V_{CKM}$.
The model has only two adjustable parameters for the PMNS mixing and 
neutrino mass ratios.  
(Other parameters are fixed from the observed quark and charged 
lepton mass ratios and the CKM mixing.) 
The model gives reasonable values $\sin^2 2\theta_{12} \simeq 0.85$,  
$\sin^2 2\theta_{23} \sim 1$ and $\sin^2 2\theta_{13} \sim 0.09$
together with $R_\nu \equiv \Delta m^2_{21}/\Delta m^2_{32} \sim 0.03$. 
Our prediction of the effective neutrino mass $\langle m \rangle$ 
in the neutrinoless double beta decay takes a sizable value
$\langle m \rangle \simeq 0.0034$ eV.
\end{abstract}

PCAC numbers:  
  11.30.Hv, 
  12.15.Ff, 
  14.60.Pq,  
  12.60.-i, 
%
\vspace{3mm}

\noindent{\large\bf 1 \ Introduction}

Many particle physicists have searched for models which provide 
a unified description of the mass spectra 
and mixing patterns of quarks and leptons, the 
Cabibbo-Kobayashi-Maskawa mixing matrix $V_{CKM}$  \cite{CKM} 
and the Pontecorvo-Maki-Nakagawa-Sakata mixing matrix 
$U_{PMNS}$ \cite{PMNS}.
As one of such models, the so-called ``yukawaon" model 
\cite{O3_PLB09,yukawaon,K-N_EPJC12,K-N_PLB12}
has been proposed.
The model is a kind of ``flavon" model \cite{flavon}. 

In this model, Yukawa coupling constants $Y_f$ ($f=u,d,e, \cdots$) 
in the standard model are understood as vacuum expectation values 
(VEVs) of scalars (``yukawaon") with $3\times 3$ components, 
i.e. by $y_f \langle Y_f \rangle /\Lambda$, where $\Lambda$ is 
an energy scale of the effective theory. 
Our policy in building the yukawaon model is as follows: 
(i) We consider that the hierarchical structures of 
the effective Yukawa coupling constants can be understood
only based on the charged lepton masses.
For the moment, we do not ask for the origin of the charged lepton 
mass spectrum.
(For an attempt to understand the origin of the charged lepton
mass spectrum, for example, see Ref.\cite{e-mass}.)
(ii) We assume a U(3) (or O(3)) family symmetry and
$R$ charge conservation. 
Structures of yukawaon VEVs $\langle Y_f \rangle$ are
obtained from SUSY vacuum conditions for a given 
superpotential, so that the VEV matrices are related to
other yukawaon VEVs. 
(As stated in (i), the charged lepton mass values are 
inputs for the moment, we do not discuss a mechanism
which gives the observed charged lepton masses.)
The first task in the yukawaon model is to search a 
superpotential form which gives reasonable mass spectra
and mixings (in other words, to search for fields with 
suitable representations of U(3) and $R$ charges.
(iii) Effect of SUSY breaking depends on a SUSY breaking
scenario.
For the moment, we do not consider the SUSY breaking
effects for yukawaon sector. 
We assume that the SUSY breaking in the quark and lepton
sectors is induced by gauge mediation (this ``gauge" 
means the conventional SU(3)$_c\times$SU(2)$_L\times$U(1)$_Y$
symmetries).
(iv) At present, our aim is to search for a mass matrix
model which can give a reasonable fit to whole of quark 
and lepton mass ratios and $V_{CKM}$ and $U_{PMNS}$ mixing matrices
with parameters as few as possible.
At present, our concern is in the construction of phenomenological mass matrix 
relations, not of a field theoretical model, i.e.
neither in economizing of the yukawaon fields nor in making 
the superpotential compact. 
It is our next step to search for a model with more economical
fields and with concise structure of superpotential. 

The yukawaon model is in the process of 
research and development at present.
In the yukawaon model, there are, in principle, no family-number-dependent 
parameters except for the charged lepton mass matrix $M_e$. 
Regrettably at present, we need a phase matrix $P_u$ (or $P_d$)
with two phase parameters in order to obtain reasonable values of 
quark mixing matrix
$V_{CKM}$ \cite{K-N_EPJC12,K-N_PLB12}. However, 
the final goal of our model is to remove such family dependent
 parameters. 
 
The yukawaon model is constructed by using fundamental VEV 
matrices of scalar fields.  
In earlier yukawaon models \cite{O3_PLB09}, the mass matrices are 
directly related to a fundamental VEV matrix matrix $\Phi_e 
\equiv {\rm diag}( \sqrt{m_e}, \sqrt{m_\mu} , \sqrt{m_\tau})$,
while in recent yukawaon models, even the charged lepton mass
matrix $M_e$ is given by a more fundamental VEV matrix $\Phi_0$.
Here, we define VEV matrices which are associated 
with the mass matrix for up, down quarks, and charged leptons 
by a common form
$$
\Phi_f = k_f \Phi_0 ( {\bf 1} + a_f X_3 ) \Phi_0 ,
\eqno(1.1)
$$ 
where $f=u,d,e$.
Here, for convenience, we have dropped the notations ``$\langle$" 
and ``$\rangle$" on the VEV matrices. 
We will assign $\Phi_0$ to $({\bf 3}^*, {\bf 3})$ 
of U(3)$\times$U(3)$'$ in the next section, so that we will denote 
$\Phi_0$ as $\bar{\Phi}_0$.
In the present section in which we discuss the VEV matrices, 
for simplicity, we do not distinguish between $\Phi_0$ and $\bar{\Phi}_0$
( and also between $Y_f$ and $\bar{Y}_f$, and so on). 
$X_3$ and ${\bf 1}$ are also VEV matrices of other scalar fields. 
The matrices $\Phi_0$, $X_3$ and ${\bf 1}$ are defined by
$$
\Phi_0 = \left( 
\begin{array}{ccc}
x_1 & 0 & 0 \\
0 & x_2 & 0 \\
0 & 0 & x_3 
\end{array} \right) , \ \ \ \ 
X_3 = \frac{1}{3} \left(
\begin{array}{ccc}
1 & 1 & 1 \\
1 & 1 & 1 \\
1 & 1 & 1 
\end{array} \right) , \ \ \ \ 
{\bf 1} = \left(
\begin{array}{ccc}
1 & 0 & 0 \\
0 & 1 & 0 \\
0 & 0 & 1 
\end{array} \right) .
\eqno(1.2)
$$
Here, we have assumed that there is a basis in which 
the VEV matrix $\Phi_0$ takes a diagonal form and the
VEV matrix $X_3$ takes a democratic form.
Our mass matrix model is described on the premise that 
there can be such the flavor basis.
The values of $(x_1, x_2, x_3)$  with $x_1^2+x_2^2+x_3^2=1$ 
are fixed by the observed charged lepton mass values 
under the given value of $a_e$.
The form $( {\bf 1} + a_e X_3 )$ is due to a family symmetry breaking
U(3)$\rightarrow$ S$_3$ \cite{K-N_PLB12} as we discuss later.
The coefficients $a_f$ play an essential role in obtaining 
the mass ratios and mixings, 
while the family-number independent coefficients $k_f$ do not. 

In this paper we propose a new model which improves the neutrino mass matrix.
As far as mass matrices $M_e$, $M_d$ and $M_u$ of the charged leptons and down- and
up-quarks are concerned, we assume the same VEV structures as those in the 
previous yukawaon model \cite{yukawaon,K-N_EPJC12,K-N_PLB12}: 
$$
M_e = \Phi_e, \ \ \ \  M_d = \Phi_d, \ \ \ \ M_u =\Phi_u \Phi_u .
\eqno(1.3)
$$
(Such the form $M_u =\Phi_u \Phi_u$ was suggested by 
a phenomenological fact 
$M_u^{diag} \sim (M_d^{diag})^2$).              
Here and hereafter, we omit family-number independent coefficients
($k_f$ in Eq.(1.1) and so on), because 
we are interested only in family structures of $3\times 3$ matrices. 
What is new in the present model is in the neutrino mass matrix $M_\nu$: 
we assume that $M_\nu$ takes the following form 
$$
M_\nu = \Phi_\nu  \Phi_\nu ,
\eqno(1.4)
$$ 
which is motivated by the up-quark mass matrix form 
$M_u = \Phi_u \Phi_u$ given in Eq.~(1.3) and 
by the correspondence
between quark and lepton mass matrices
$M_e \leftrightarrow M_d$ and $M_\nu \leftrightarrow M_u$. 
The newly introduced VEV matrix $\Phi_\nu$ in Eq.~(1.4) is given by
$$  
\Phi_\nu= M_D M_R^{-1} M_D^T .
\eqno(1.5)
$$
Here we take 
$$
M_D = \Phi_D = \Phi_0^T ( {\bf 1} + a_D X_2 ) \Phi_0 ,
\eqno(1.6)
$$
$$
M_R = \Phi_u \Phi_e + \Phi_e \Phi_u  . 
\eqno(1.7)
$$
Though we use the notations  $M_D$ and $M_R$ in Eqs.(1.5) - (1.7) , 
they have no meaning of the Dirac or the right handed Majorana neutrino 
mass matrices  differently from the previous model [see Eq.(2.3) later].
Note also that the form of $M_D$ given by Eq.~(1.6) is different from that of other 
VEV matrices given by Eq.~(1.1).
Here, the matrix form $X_2$ \cite{K-N_EPJC13} is defined by
$$
X_2 = \frac{1}{2} \left(
\begin{array}{ccc}
1 & 1 & 0 \\
1 & 1 & 0 \\
0 & 0 & 0 
\end{array} \right),
\eqno(1.8)
$$
which will be discussed in Section 2.

Let us stress the difference of the form for the neutrino mass matrix 
between the present model and the previous one. In the previous yukawaon model
\cite{yukawaon,K-N_EPJC12,K-N_PLB12}, the neutrino mass 
matrix $M_\nu$ was given by a form
$$
\begin{array}{l}
M_\nu = M_D M_R^{-1} M_D^T , \\
M_D = M_e , \\
M_R = (\Phi_u M_e + M_e \Phi_u ) + \xi_\nu \ 
{\rm term}, 
\end{array}
\eqno(1.9)
$$
where $\xi_\nu$-term was an additional term which 
was brought in order to fit neutrino mixing
parameters $\sin^2 \theta_{23}$ and $\sin^2 \theta_{12}$. 
However, the model could not give reasonable fit for
$\sin^2 \theta_{13}$. On the other hand, 
the mass matrix (1.4) with (1.7) in the new model 
has no such the $\xi_\nu$-term. 
Nevertheless, we can fit whole the observed 
mixing values $\sin^2 \theta_{23}$, $\sin^2 \theta_{12}$ 
and $\sin^2 \theta_{13}$ together with the ratio of neutrino mass-squared difference
$R_\nu = \Delta m^2_{21}/\Delta m^2_{32}$ by using (1.5), as stated in Section 3. 
(The big drawback in the previous yukawaon models 
was that the model 
could not give the observed large value \cite{theta13} of 
$\sin^2 2\theta_{13}\sim 0.09$.)

In Sec.2, we give VEV matrix relations in the new model. 
In Sec.3, we discuss parameter fitting of observed values 
only for the PMNS mixing and neutrino mass ratios 
because we revised the model only in the neutrino sector.
The parameter values in the down-quark sector are effectively 
unchanged, so that we can obtain the same predictions for the down-quark mass 
ratios and CKM matrix parameters without changing the successful results 
in the previous paper \cite{K-N_EPJC13}.

\vspace{3mm}

\noindent{\large\bf 2 \ VEV matrix relations}

We assume that a would-be Yukawa interaction is given as follows:
$$
W_Y = \frac{y_e}{\Lambda} e^c_i \bar{Y}_e^{ij} \ell_j H_d 
+ \frac{y_\nu}{\Lambda^2} (\ell_i H_u) \bar{Y}_\nu^{ij} (\ell_j H_u)
+ \frac{y_d}{\Lambda} d^{ci} Y^d_{ij} q^j H_d  
+ \frac{y_u}{\Lambda} u^{ci} Y^u_{ij} q^j H_u  ,
\eqno(2.1)
$$
where $\ell=(\nu_L, e_L)$ and $q=(u_L, d_L)$ are SU(2)$_L$ doublets. 
Assignments of these fields to family symmetries U(3)$\times$U(3)$'$
are given in Table 1. 
We denote the yukawaons with $({\bf 6}^*, {\bf 1})$ and
$({\bf 6}, {\bf 1})$ as $\bar{Y}$ and $Y$, respectively.
Note that in Eq.(2.1) there are no SU(2)$_L$ singlet neutrinos.
We have straightforwardly defined the neutrino mass matrix $M_\nu$
by the second term in Eq.(2.1). 
Although we denoted in Eq.(1.6) as if the matrix $M_D$ is a Dirac
neutrino mass matrix, the matrix $M_D$ does not have a meaning of
the Dirac mass matrix [see Eq.(2.3) later].
Under the definition of $\bar{Y}_\ell$ ($Y^q$) in Eq.(2.1), 
the quark mixing matrix $V_{CKM}$ and 
the lepton mixing mixing matrix $U_{PMNS}$ are given by 
$V_{CKM}=U_u^\dagger U_d$ 
and $U_{PMNS}=U_e^\dagger U_\nu$, respectively, where
$U_f$ are defined by $U_f^\dagger M_f^\dagger M_f U_f
= D_f^2$ ($D_f$ are diagonal).  
Here and hereafter, sometimes, we denote $\bar{Y}_\ell$ 
and $Y^q$ as $Y_f$ for simplify.
In order to distinguish each yukawaon from others, we assume that 
$Y_f$ have different $R$ charges from each other under consideration of $R$ 
charge conservation.
(Of course, the $R$ charge conservation is broken
at the energy scale $\Lambda$.)

\begin{table}[h]
\caption{ SU(2)$_L \times$SU(3)$_c \times$U(3)$
\times$U(3)$'$ assignments and $R$ charges}

\vspace{2mm}
\begin{center}
\begin{tabular}{|c|ccccc|cc|cccc|} \hline
& $\ell$ & $e^c$  & $q$ & $u^c$ & $d^c$ & $H_u$ & $H_d$ &
$\bar{Y}_e$ & $\bar{Y}_\nu$ & $Y^d$ & $Y^u$ \\ \hline
SU(2)$_L$ & ${\bf 2}$ & ${\bf 1}$ & ${\bf 2}$ & ${\bf 1}$ & 
${\bf 1}$ & ${\bf 2}$ & ${\bf 2}$  & 
${\bf 1}$ & ${\bf 1}$ & ${\bf 1}$ & ${\bf 1}$ \\
SU(3)$_c$ & ${\bf 1}$ & ${\bf 1}$ & ${\bf 3}$ & 
${\bf 3}^*$ & ${\bf 3}^*$ & ${\bf 1}$ & ${\bf 1}$ & 
${\bf 1}$ & ${\bf 1}$ & ${\bf 1}$ & ${\bf 1}$ \\
U(3) & ${\bf 3}$ & ${\bf 3}$  & ${\bf 3}^*$ & 
${\bf 3}^*$ & ${\bf 3}^*$ & ${\bf 1}$ & ${\bf 1}$ &
${\bf 6}^*$ & ${\bf 6}^*$ & ${\bf 6}$ & ${\bf 6}$   \\
U(3)$'$ & ${\bf 1}$ & ${\bf 1}$ & ${\bf 1}$ & 
${\bf 1}$ & ${\bf 1}$ & ${\bf 1}$ & ${\bf 1}$ &
${\bf 1}$ & ${\bf 1}$ & ${\bf 1}$ & ${\bf 1}$  \\ 
$R$ & $r_\ell$ & $r_{ec}$ & $r_q$ & $r_{uc}$ & $r_{dc}$ & 
$r_{Hu}$ & $r_{Hd}$ &
$\bar{r}_{Ye}$ & $\bar{r}_{Y\nu}$ & $r_{Yd}$ & $r_{Yu}$ \\ \hline
\end{tabular}
\end{center}
\end{table}

We obtain VEV matrix relations from the
superpotential which is invariant under the family 
symmetries U(3)$\times$U(3)$'$ and is $R$ charge conserving. 
In the yukawaon model, the VEV matrix relations are 
phenomenological ones, and they are dependent on the $R$ 
charge assignments. 
Since derivations of the VEV matrix relations are essentially
similar to those in the previous papers 
\cite{O3_PLB09,yukawaon,K-N_EPJC12,K-N_PLB12,K-N_EPJC13}, 
although the  U(3)$\times$U(3)$'$ assignments  and $R$ charges 
are different.  
Besides, we must consider a complicated superpotential form 
in order to derive the desirable mass matrix relations.
The purpose of the present paper is not to derive those mass matrix
relations uniquely, but to investigate a possibility that 
the neutrino mass 
matrix $M_\nu$ is given by a form $M_\nu = (M_D M_R^{-1} M_D^T)^2$, from the 
phenomenological point of view.
Therefore, in this section, we present only the results of 
the mass matrix relations, 
the derivation of which is discussed in Appendix: 
$$
\langle \bar{Y}_e \rangle = \langle \bar{\Phi}_e \rangle = 
\langle \bar{\Phi}_0 \rangle \left( {\bf 1} + 
a_e X X^T \right) \langle \bar{\Phi}_0^T \rangle ,
\eqno(2.2)
$$
$$
 \langle \bar{Y}_\nu \rangle  
 = \langle \bar{\Phi}_\nu \rangle  \langle \bar{\Phi}_\nu \rangle ,
 \eqno(2.3)
$$
$$
 \langle \bar{\Phi}_\nu \rangle =
 \langle \bar{Y}_D \rangle 
(\langle \bar{Y}_R \rangle)^{-1}  \langle \bar{Y}_D \rangle, 
 \eqno(2.4)
$$ 
$$
\langle \bar{Y}_D \rangle = \langle \bar{\Phi}_0^T \rangle \left( {\bf 1} + 
a_D X^T X \right) \langle \bar{\Phi}_0 \rangle ,
\eqno(2.5)
$$
$$
\langle \bar{Y}_R \rangle =  \langle \bar{Y}_e \rangle 
\langle \Phi^u \rangle + \langle \Phi^u \rangle
\langle \bar{Y}_e \rangle ,
\eqno(2.6)
$$
$$
 \langle {Y}^u \rangle  
 =  \langle \Phi^u \rangle \langle \Phi^u \rangle ,
 \eqno(2.7)
$$
$$
\langle \Phi^u \rangle =  \langle \bar{\Phi}_0 \rangle \left( {\bf 1} + 
a_u X X^T \right) \langle \bar{\Phi}_0^T \rangle ,
\eqno(2.8)
$$
$$
\langle \bar{P}_d \rangle  \langle Y^d \rangle  
\langle \bar{P}_d \rangle  =
  \langle \bar{\Phi}_0 \rangle \left( {\bf 1} + 
a_d X X^T \right) \langle \bar{\Phi}_0^T \rangle  + \xi_0^d {\bf 1} .
\eqno(2.9)
$$
Here, the fields $\bar{\Phi}_0^{i\alpha}$ and $X_{\alpha i}$ are assigned to
$({\bf 3}^*, {\bf 3}^*)$ and $({\bf 3}, {\bf 3})$ of U(3)$\times$U(3)$^\prime$, 
respectively.  
The field $X$ has phenomenologically been introduced in the previous model 
\cite{K-N_EPJC13}, the VEV of which has the form
$$
\frac{1}{v_X} \langle X \rangle_{\alpha i} = \frac{1}{2} \left(
\begin{array}{ccc}
1 & 1 & 0 \\
1 & 1 & 0 \\
1 & 1 & 0 
\end{array} \right)_{\alpha i} .
\eqno(2.10)
$$
The form (2.10) leads to
$$
\left(\langle X \rangle \langle X^T \rangle\right)_{\alpha\beta} = 
\frac{3}{2} (X_3)_{\alpha\beta}, \ \ \ \ 
\left(\langle X^T \rangle \langle X \rangle \right)_{ij} = 
\frac{3}{2} (X_2)_{ij}, 
\eqno(2.11)
$$
together with $\langle X \rangle \langle X \rangle=\langle X \rangle$, 
where $X_3$ and $X_2$ is defined by 
Eqs. (1.2) and (1.8), respectively.
Here, for simplicity, 
we have put $v_X=1$ because we are interested only in the relative 
ratios among the family components.      
At present, there is no idea for the origin of the form (2.10). 
We may speculate that this form is related to a breaking pattern
of U(3)$\times$U(3)$'$ (for example, 
discrete symmetries U(3)$\times$U(3)$' \rightarrow$S$_2 \times$S$_3$). 
In the present paper, the form (2.10) is only ad hoc assumption. 
However, as seen later, we can obtain a good fitting for the neutrino 
mixing angle $\sin^2 2 \theta_{13}$ due to this assumption.

\vspace{3mm}

\noindent{\large\bf 3 \ Parameter fitting}

We again summarize our mass matrix model as follows:  
$$
M_e= \bar{Y}_e  =  \bar{\Phi}_0 ( {\bf 1} + a_e X_3 ) \bar{\Phi}_0^T ,
\eqno(3.1)
$$
$$
M_\nu =\bar{Y}_\nu = \bar{\Phi}_\nu \bar{\Phi}_\nu ,
\eqno(3.2)
$$
$$
\Phi_\nu = \bar{Y}_D \bar{Y}_R^{-1} \bar{Y}_D ,
\eqno(3.3)
$$
$$
M_D = \bar{Y}_D =  \bar{\Phi}_0^T ( {\bf 1} + a_D e^{i \alpha_D} X_2) \bar{\Phi}_0 ,
\eqno(3.4) 
$$
$$
M_R= \bar{Y}_R =  \left( \bar{Y}_e \Phi^u + \Phi^u \bar{Y}_e  \right),
\eqno(3.5)
$$
$$
M_u = Y^u =  \Phi^u \Phi^u ,
 \eqno(3.6)
$$
$$
\Phi^u =  \bar{\Phi}_0  \left( {\bf 1} + 
a_u e^{i \alpha_u} X_3 \right) \bar{\Phi}_0^T ,
\eqno(3.7)
$$
$$
\bar{P}_d  Y^d \bar{P}_d =  \bar{\Phi}_0 ( {\bf 1} + a_d X_3) 
\bar{\Phi}_0^T +\xi_0^d {\bf 1} ,
\eqno(3.8)
$$
where, for convenience, we have dropped 
the notations ``$\langle$" and ``$\rangle$". 
In numerical calculations,  
we use  dimensionless expressions
$\bar{\Phi}_0 = {\rm diag}(x_1, x_2, x_3)$ 
(with $x_1^2+x_2^2+x_3^2=1$) and  
$\bar{P}_d= {\rm diag} (e^{-i\phi_1}, e^{-i\phi_2},1)$. 
The parameters are re-refined by Eqs.(3.1)-(3.8). 
In Eqs.(3.7) and (3.4), we have denoted $a_u$ and $a_D$ as 
$a_u e^{i \alpha_u}$ and 
$a_D e^{i \alpha_D}$, respectively, since we assume that the parameters 
$a_e$ and $a_d$ are real, while $a_u$ and $a_D$ are complex
in our $M_D\leftrightarrow M_u$ and $M_e\leftrightarrow M_d$  
correspondence scheme.

In this model, we have two parameters $(a_D, \alpha_D)$ for 
neutrino sector, four parameters $a_D$, $\xi_0^d$ and 
$(\phi_1, \phi_2)$ for down-quark mass ratios and 
$V_{CKM}$, and three parameters $a_e$, $(a_u, \alpha_u)$
for charged lepton mass ratios and up-quark mass ratios
as shown in Table 2.
Especially, it is worthwhile noticing that the neutrino 
mass ratios and $U_{PMNS}$ are functions of only two parameters 
after $a_e$ and $(a_u, \alpha_u)$ have been fixed from the 
observed CKM mixing and up-quark mass ratios.
There is effectively no change in the mass matrix structures 
except for $Y_\nu$ from the previous paper \cite{K-N_EPJC13}, so that
we can use the same parameter values for $a_e$ and $(a_u, \alpha_u)$ 
as those in the previous study \cite{K-N_EPJC13}, which are given by 
$$
a_e =7.5, \ \ \  (a_u, \alpha_u) = ( -1.35, 7.6^\circ) .
\eqno(3.9)
$$ 
Therefore, as far as PMNS mixing and neutrino mass ratios
are concerned,  
we have only two free parameters $(a_D, \alpha_D)$
in the present neutrino mass matrix model.  

\begin{table}
\caption{Process for parameter fitting. 
Since the parameters listed in each step can slightly 
affect predictions listed in the other steps, we need
fine tuning after the 5h step.  
New parameter fitting in the present paper starts  
from the 5th step.
}
\vspace{2mm}
\begin{center}
\begin{tabular}{|c|cc|cc|c|} \hline
Step & Inputs & $N_{inp}$ &  Parameters & $N_{par}$ &
 Predictions  \\ \hline
1st &  $\frac{m_e}{m_\mu}$, $\frac{m_\mu}{m_\tau}$  
& 4 & $\frac{x_1}{x_2}$, $\frac{x_2}{x_3}$ & 4 &  \\
  & $\frac{m_u}{m_c}$, $\frac{m_c}{m_t}$ & & $a_e$, $a_u$ &  &    
\\ \hline
2nd  & $|V_{us}|$, $|V_{cb}|$, $|V_{ub}|$ & 3 & $\alpha_u$, 
$(\phi_1, \phi_2)$ &  3 &  $|V_{td}|$, $\delta_{CP}^q$ \\ \hline
3rd  & $\frac{m_s}{m_b}$ & 1 & $a_d$ &  1 &   \\ \hline
4th  & $\frac{m_d}{m_s}$ & 1 & $m_d^0$ & 1 &   
not affect to other predictions  \\ \hline 
5th  & $\sin^2 2\theta_{12}$  & 2 & $a_D$ & 2 &
$\sin^2 2\theta_{13}$, $\delta_{CP}^\ell$, 2 Majorana phases \\
    &   $R_\nu$ &   & $\alpha_D$  &  & $\sin^2 2\theta_{23}$, 
$\frac{m_{\nu 1}}{m_{\nu 2}}$,
 $\frac{m_{\nu 2}}{m_{\nu 3}}$   \\ \hline
option &  $\Delta m^2_{atm}$ &  1 & $m_{\nu 3}$ & 1 & 
$(m_{\nu 1}, m_{\nu 2},  m_{\nu 3})$, $\langle m \rangle$  \\
\hline
$\sum N_{\dots}$ & & 12 &  & 12 &   \\ \hline 
\end{tabular}
\end{center}
\end{table}

At present, the observed values \cite{PDG2012} are as follows:
$$
\sin^2 2\theta_{12}^{obs}= 0.857 \pm 0.024, \ \ \ 
\sin^2 2\theta_{23}^{obs} > 0.95, \ \ \ 
\sin^2 2\theta_{13}^{obs}= 0.098 \pm 0.013, 
\eqno(3.10)
$$
$$
R_{\nu}^{obs} \equiv \frac{(\Delta m_{21}^2)^{obs}}{
(\Delta m_{32}^2)^{obs}}=
\frac{(7.50\pm 0.20) \times 10^{-5}\ {\rm eV}^2}{
(2.32^{+0.12}_{-0.08}) \times 10^{-3}\ {\rm eV}^2} = 
(3.23^{+0.14}_{-0.19} ) \times 10^{-2} .
\eqno(3.11)
$$

Since the parameters $(a_D, \alpha_D)$ are sensitive to the 
observables $\sin^2 2\theta_{12}^{obs}$ and $R_\nu^{obs}$, we use 
the observed values of $\sin^2 2\theta_{12}$ and $R_\nu$  
in order to fix our parameter values $(a_D, \alpha_D)$.
In Fig.1, we illustrate an allowed parameter region of $(a_D, \alpha_D)$
obtained from the observed values of 
$\sin^2 2\theta_{12}^{obs}$ and $R_\nu^{obs}$. 
As seen in Fig.1, the observed values uniquely fix the 
parameter values  $(a_D, \alpha_D)$ as
$$
(a_D, \alpha_D) = (8.7, 12^\circ) .
\eqno(3.12)
$$
It is worthwhile noticing that the parameter values (3.12)
uniquely give a prediction of $\sin^2 2\theta_{13} \simeq 0.09$. 
For reference, in Fig.2,  we illustrate behaviors of $\sin^2 2\theta_{12}$ 
and $R_\nu$ versus $\alpha_D$ in the case of $a_D=8.7$. 
We find that the choice $\alpha_D=12^\circ$ gives excellent fittings to the
observed values of $\sin^2 2\theta_{12}$ and $R_\nu$ simultaneously: 
$$
\sin^2 2\theta_{12}= 0.8544, \ \ \ R_\nu = 0.0331 .
\eqno(3.13)
$$
Then, we obtain our predictions for $\sin^2 2\theta_{23}$ and 
$\sin^2 2\theta_{13}$ using (3.12) as follows: 
$$
\sin^2 2\theta_{23}= 0.9962, \ \ \ 
\sin^2 2\theta_{13}= 0.0907 ,  
\eqno(3.14)
$$
which are in excellent agreement with the observed values given
in Eq.(3.10).


\begin{figure}[h]
\begin{picture}(200,200)(0,0)

\includegraphics[height=.3\textheight]{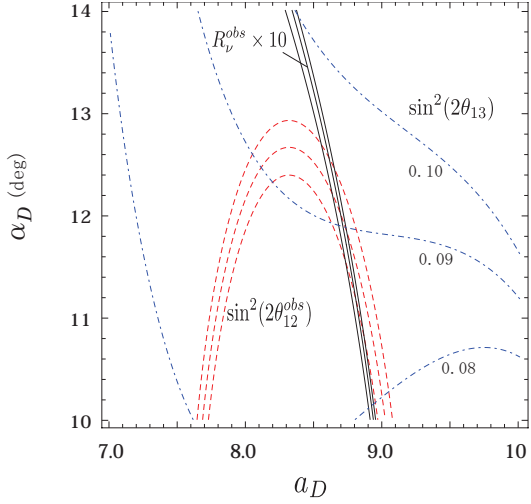}
\end{picture}  
  \caption{Allowed parameter region in $(a_D, \alpha_D)$ plane. 
The solid and dashed curves indicate the border and center curves 
of the allowed region which are obtained from 
the observe values of $\sin^2 2\theta_{12}^{obs}$ and 
$R_\nu^{obs}\times 10$, respectively.
The dot-dashed curves represent contour curves of $\sin^2 2\theta_{13}$ for some typical values, 
$\sin^2 2\theta_{13}=0.08$, $0.09$, and $0.10$. }
  \label{fig1}
\end{figure}

\begin{figure}[h]
\begin{picture}(200,200)(0,0)

\includegraphics[height=.3\textheight]{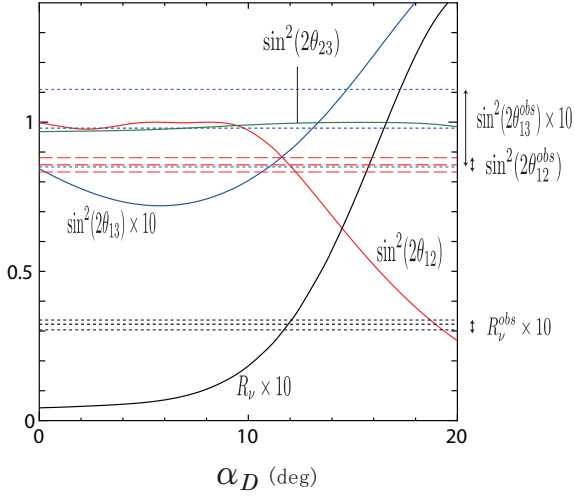}
\end{picture}  
  \caption{{Lepton mixing parameters $\sin^2 2\theta_{12}$, 
$\sin^2 2\theta_{23}$, $\sin^2 2\theta_{13}$, and the ratio 
$R_\nu$ versus the phase parameter $\alpha_D$
for $a_D=8.7$. 
The horizontal lines denote observed values (the center
and 1$\sigma$ values) of $\sin^2 2\theta_{12}^{obs}$,
$\sin^2 2\theta_{13}^{obs}\times 10$ and
$R_\nu^{obs}\times 10$. 
Our predicted value for $\sin^2 2\theta_{23}$ is well 
satisfied the obtained experimental bound of 
$\sin^2 2\theta_{23}^{obs}$.
}
}
  \label{fig2}
\end{figure}

The fixing of the parameters  $(a_D, \alpha_D)$, Eq.(3.12), leads to the 
prediction of the $CP$ violating phase parameter in the lepton sector too: 
$$
\delta_{CP}^{\ell}= 127^\circ  \ \ \ (J^{\ell} = 2.74 \times 10^{-2} ), 
\eqno(3.15)
$$
where $\delta_{CP}^{\ell}$ is the $CP$ violating phase in the standard 
expression and $J^{\ell}$ is the rephasing invariant \cite{J}.
We can also predict neutrino masses:  
$$
m_{\nu 1} = 0.00061\ {\rm eV}, \ \ m_{\nu 2} = 0.00899 \ {\rm eV}, 
\ \ m_{\nu 3} = 0.05011\ {\rm eV}  ,
\eqno(3.16)
$$
by using the input value \cite{MINOS08}
$\Delta m^2_{32} = 0.00243$ eV$^2$. 
(Note that, in the present model, we cannot obtain an inverted 
neutrino mass hierarchy, because the hierarchies of the mass matrices
are related to the hierarchy of the charged lepton mass hierarchy, 
i.e. to the VEV matrix $\langle \Phi_0 \rangle$.)
We also predict the effective Majorana neutrino mass \cite{Doi1981} 
$\langle m \rangle$ 
in the neutrinoless double beta decay as
$$
\langle m \rangle =\left|m_1 U_{e1}^2 +m_2 U_{e2}^2 
+m_3 U_{e3}^2\right| = 0.0034 \ {\rm eV}.
\eqno(3.17)
$$
This predicted value is considerably larger than those in 
other models with normal hierarchy
\cite{bb_expmt}. 

Finally, we list the predicted values of the CKM mixing parameters 
and down-quark mass ratios, although they are essentially the same 
as those in the previous model \cite{K-N_EPJC13}:
$$
|V_{us}|=0.2271, \ \ \ |V_{cb}|=0.0394, \ \ \ 
|V_{ub}|=0.00347, \ \ \ |V_{td}|= 0.00780 ,
\eqno(3.18)
$$
$$
\delta_{CP}^{q}= 59.6^\circ   \ \ \ (J^{q} = 2.6 \times 10^{-5} ) , 
\eqno(3.19)
$$
$$
r^u_{12} = \sqrt{\frac{m_d}{m_s}}= 0.00465, \ \ \ \ 
r^u_{23} = \sqrt{\frac{m_d}{m_b}}= 0.0614 .
\eqno(3.20)
$$
$$
r^d_{12} = \frac{m_d}{m_s}= 0.0569, \ \ \ \ 
r^d_{23} = \frac{m_d}{m_b}= 0.0240 .
\eqno(3.21)
$$
Here, we have used $a_d = 25$,  $\xi_0^d= 0.0115$, and 
$(\phi_1, \phi_2)=(177.0^\circ, 197.4^\circ)$. 
The observed values are as follows:
$|V_{us}|=0.2252 \pm 0.0009$, $|V_{cb}|=0.0409 \pm 0.0011$,
$|V_{ub}|=0.00415 \pm 0.00049$, $|V_{td}|= 0.0084 \pm 0.0006$, 
$J^q=(2.96^{+0.20}_{-0.16}) \times 10^{-5}$ \cite{PDG2012}, and 
$r^u_{12} =0.045^{+0.013}_{-0.010}$, $r^u_{23}=0.060\pm 0.005$, 
$r^d_{12}=0.053^{+0.005}_{-0.003}$, $r^d_{23}=0.019 \pm 0.006$ 
 \cite{q-mass}.

\vspace{3mm}

\noindent{\large\bf 4 \ Concluding remarks}

In conclusion, we have proposed a new neutrino mass 
matrix form within the framework of the yukawaon model, 
in which we have only two adjustable parameters, $(a_D, \alpha_D)$, 
for PMNS mixing and neutrino mass ratios. 
We have been able to remove the unnatural term 
[$\xi_\nu$ term in Eq.(1.9)] in the previous model. 
Nevertheless, we can obtain reasonable results for PMNS mixing 
and neutrino mass ratios as shown in Eqs.(3.13) - (3.17)
for the parameter values $(a_D, \alpha_D)=(8.7, 12^\circ)$. 
As seen in Fig.2, it is worthwhile noticing that 
only when we choose a reasonable value of $R_\nu \simeq 0.033$, 
we can obtain a reasonable value of $\sin^2 2\theta_{13}
\simeq 0.09$.
Also, note that our prediction
gives a sizable value of $\langle m\rangle \simeq 0.0034$ eV
among normal mass hierarchy models. 
Of course, we have also obtained reasonable results 
for CKM mixing and quark mass ratios as same as 
those in the previous paper \cite{K-N_EPJC13}.

Such the phenomenological success is essentially based
on the following assumptions:
(i) We have assumed that only $Y_D$ takes the 
mass matrix form with $X_2$ (not $X_3$), while 
others $Y_f$ ($\Phi_f$) take the form with $X_3$
as given in Eq.(1.1). 
In Ref.\cite{K-N_PLB12}, the form $X_3$ has been 
understood by a symmetry breakdown 
U(3)$\times$U(3)$' \rightarrow$ U(3)$\times$S$_3$. 
However, for the form $X_2$, the model is still 
in a phenomenological level.  
(ii) We have the bilinear form of the neutrino
mass matrix, $M_\nu = \Phi_\nu \Phi_\nu$, as 
well as the up-quark mass matrix $M_u = \Phi_u \Phi_u$.
From the theoretical point view, there is no reason 
for the bilinear forms.
We merely assigned $R$ charges so that bilinear forms 
are realized for $M_u$ and $M_\nu$.

In spite of such the phenomenological success,  
the model still leave some basic problems:
(i) The model is not economical.
At present, we need many flavons in order to prepare 
reasonable VEV matrix relations. 
Since the purpose of the present paper is to 
investigate phenomenological relations among
mass matrices, 
the structure of the superpotential given in Appendix is a temporal one.  
The superpotential will be improved in our future work.
(ii) We have not discuss scales of yukawaons.
The present model is based on an effective theory with 
an energy scale $\Lambda$.
The scale $\Lambda$ must be, at least, larger than
$10^3$ TeV from the observed $K^0$-$\bar{K}^0$ mixing
(and also $D^0$-$\bar{D}^0$ mixing) \cite{PDG2012}.
In earlier version of the yukawaon model, it was 
considered to be $\Lambda \sim 10^{15}$ GeV.
However, VEVs of individual yukawaons depend on
parameters in the superpotential ($\mu_f$ in mass terms 
and couplings $\lambda_f$). 
We do not fix those scales in the present paper, 
although we expect that effects of those flavons are visible. 
(iii) We did not discuss SUSY breaking effects. 
As we stated in Section 1, for the time being, we 
assume that the SUSY breaking effects do not affect
yukawaon sector. 
(iv) Our goal is to understand the hierarchical structures
of all quark and lepton mass matrices on the basis of 
only the observed charged lepton masses. 
However, in the present model, we are still obliged to
introduce flavon $\bar{P}_d$ whose VEV matrix includes
flavor-dependent parameters $\phi_1$ and $\phi_2$ as  
seen in (A.11).

Generally speaking, the yukawaon model suggests that our direction
to unified understanding of the flavor problems 
is not wrong, although we have many problems in the yukawaon model.
By leaving the settlement of the problems to our future tasks, 
the yukawaon model will be improved step by step.

\vspace{5mm}

\centerline{\Large\bf  \ Appendix} 

In this Appendix, we discuss a derivation of the mass matrix relations (2.2)-(2.9) 
from superpotential. 
We assume the following  superpotential $W=W_e+W_\nu + W_R+ W_D +
W_u + W_d$:  
$$
W_e=  \left\{ \mu_e \bar{Y}_e^{ij} + \frac{\lambda_e}{\Lambda} 
(\bar{\Phi}_0)^{i\alpha}
 \left( E^{\prime\prime}_{\alpha\beta} 
+  \frac{a_e}{\Lambda^2} X_{\alpha k} \bar{E}^{kl} X^T_{l\beta} \right)
 (\bar{\Phi}_0^T)^{\beta j} \right\} \Theta^e_{ji} ,
\eqno(A.1)
$$
$$
W_\nu = \frac{1}{\Lambda}  \left[ {\lambda_\nu} (E')^\alpha_k 
\bar{Y}_\nu^{kl} (E^{\prime T})^\beta_l 
+{\lambda'_\nu} (\Phi_\nu^T)^{\alpha\gamma}
E^{\prime\prime}_{\gamma\delta} \Phi_\nu^{\delta\beta} \right] 
\Theta_{\beta\alpha}^\nu
+ \left[ \mu_\nu \Phi_\nu^{\alpha\beta} + \frac{\hat{\lambda}_\nu}{\Lambda} 
\bar{Y}_D^{\alpha\gamma} \Phi^R_{\gamma\delta}  \bar{Y}_D^{\delta\beta}
 \right] \hat{\Theta}^{\nu}_{\beta\alpha} ,
\eqno(A.2)
$$
$$
W_D =\left[ \mu_D \bar{Y}_D^{\alpha\beta}  
+\frac{\lambda_D}{\Lambda} (\bar{\Phi}_0^T)^{\alpha k}
\left( E_{kl} +\frac{\lambda_D^{\prime}}{\Lambda^2} X^T_{k\beta} 
(\bar{E}^{\prime\prime})^{\beta\gamma}
X_{\gamma l} \right) \bar{\Phi}_0^{l\beta} \right] \Theta^D_{\beta\alpha} ,
\eqno(A.3)
$$
$$
W_R=  \left[ \frac{\lambda_R}{\Lambda} \bar{Y}_R^{ik} (E')^\gamma_k
\Phi^R_{\gamma \alpha} 
+ \mu_R (\bar{E}')^i_\alpha \right] (\hat{\Theta}_R)^\alpha_i 
+ \left[ \mu_R \bar{Y}_R^{i j} + \frac{\lambda'_R}{\Lambda} \left(
\bar{Y}_e^{ik} \Phi^u_{kl} \bar{E}^{lj} +
\bar{E}^{ik} \Phi^u_{kl} \bar{Y}_e^{lj} 
\right) \right] \Theta^R_{ji} ,
\eqno(A.4)
$$
$$
W_u=   \left( \mu_u  Y^u_{ij} +\frac{\lambda_u }{\Lambda} 
\Phi^u_{ik} \bar{E}^{kl}
 \Phi^u_{lj}\right) \bar{\Theta}_u^{ji} 
$$
$$
+ \frac{1}{\Lambda} \left[ \lambda'_u \bar{E}^{ik}_u \Phi^u_{kl}
\bar{E}^{lj}_u  + \lambda^{\prime\prime}_u
(\bar{\Phi}_0)^{i\alpha} \left( (E^{\prime\prime}_u)_{\alpha\beta} 
+ \frac{a_u }{\Lambda^2}
X_{\alpha k} \bar{E}^{kl}_u X^T_{l\beta} \right) 
(\bar{\Phi}_0^T)^{\beta j}\right] \hat{\Theta}^{u}_{ji}, 
\eqno(A.5)
$$
$$
W_d=  \left[ \frac{\lambda_d}{\Lambda}\lambda_d \bar{P}_d^{ik} 
{Y}^d_{kl} \bar{P}_d^{lj} +  \frac{\lambda'_d}{\Lambda}
(\bar{\Phi}_0)^{i\alpha} \left( (E^{\prime\prime}_d)_{\alpha\beta} + 
 \frac{a_d}{\Lambda^2} X_{\alpha k} \bar{E}^{kl}_d X^T_{l\beta} 
\right) (\bar{\Phi}_0^T)^{\beta j} + \mu_d \bar{E}_d^{ij}
 \right] \Theta^{d}_{ji} . 
\eqno(A.6)
$$
The VEV matrix relations (2.2) - (2.9) are obtained from
SUSY vacuum conditions, $\partial W/\partial \Theta_A =0$
($A= e, \nu, \cdots$).
Since we assume that all $\Theta$ fields take $\langle \Theta \rangle=0$,
SUSY vacuum conditions with respect to another fields do not lead to
meaningful relations, because such conditions always contain, 
at least, one  $\langle \Theta \rangle$. 

In Eqs.(A.5) and (A.6), we have introduced fields $E^{\prime\prime}_u$, 
$E^{\prime\prime}_d$, $\bar{E}_u$ and $\bar{E}_d$ 
in addition to $E^{\prime\prime}$ and $\bar{E}$ 
in order to distinguish the $R$ charges of 
$\hat{\Theta}^{u}$ and $\Theta^{d}$ from that of $\Theta^{e}$. 
All VEV matrices $\langle E \rangle$ are given by the forms 
$\langle E \rangle \propto {\bf 1}$ as seen in (A.10).
The VEV matrix relations (2.2) - (2.9) have already been presented  
by replacing $\langle E \rangle \rightarrow {\bf 1}$. 
 
We list the SU(2)$_L \times$SU(3)$_c\times$U(3)$\times$U(3)$'$ 
assignments and 
$R$ charges for additional fields in Table 3. 
The assignments of $R$ charges are done so that the total $R$ charge 
of the superpotential term is $R(W)=2$. 
We have 17 constraints on the $R$ charges of the fields from 
Eqs.(2.1) and (A.1) - (A.6), while we have 34 fields even except for 
$\Theta$ fields in Tables 1 and 3.
Therefore, we cannot uniquely fix $R$ charge assignments of those
fields.
Here, let us give only typical constraints:
$$
2 r_X = r^{\prime\prime}_E - \bar{r}_E = r_E -\bar{r}^{\prime\prime}_E
= r^{\prime\prime}_{Eu} - \bar{r}_{Eu} 
= r^{\prime\prime}_{Ed} - \bar{r}_{Ed} , 
\eqno(A.7)
$$
$$
2 r_0 = \bar{r}_{Ye} - r^{\prime\prime}_E 
= \bar{r}_{YD} - r_E 
= \hat{r}_{Yu} + 2 \bar{r}_{Eu} - r^{\prime\prime}_{Eu}
= \hat{r}_{Yd} + 2 \bar{r}_{Pd} - r^{\prime\prime}_{Ed} .
\eqno(A.8)
$$

\begin{table}[h]
\caption{SU(2)$_L \times$SU(3)$_c \times$U(3)$\times$U(3)$'$ 
assignments and $R$ charges}

\vspace{2mm}
\begin{center}
\begin{tabular}{|c|ccccc|ccccccccc|} \hline
 & 
$\Phi_\nu$ & $\bar{Y}_D$ & $\bar{Y}_R$& $\Phi^R$ & $\Phi^u$ &  
$\Theta^e$ & $\Theta^\nu$ & $\hat{\Theta}^\nu$ &  $\Theta^D$ & 
$\Theta^R$ & $\hat{\Theta}_R$ & $\bar{\Theta}_u$ & 
$\hat{\Theta}^{u}$ & $\Theta^d$ \\ \hline
SU(2)$_L$ &${\bf 1}$ & ${\bf 1}$ & ${\bf 1}$ & ${\bf 1}$ & ${\bf 1}$ & 
${\bf 1}$ & ${\bf 1}$ &${\bf 1}$ & ${\bf 1}$ & ${\bf 1}$ & ${\bf 1}$ &
${\bf 1}$ & ${\bf 1}$ & ${\bf 1}$  \\
SU(3)$_c$ &  ${\bf 1}$ & ${\bf 1}$ &  ${\bf 1}$ & ${\bf 1}$ & ${\bf 1}$ & 
${\bf 1}$ & ${\bf 1}$ & ${\bf 1}$ & ${\bf 1}$ & ${\bf 1}$ & ${\bf 1}$ &
${\bf 1}$ & ${\bf 1}$ & ${\bf 1}$  \\
SU(3) & 
${\bf 1}$ & ${\bf 1}$ & ${\bf 6}^*$ & ${\bf 1}$ & ${\bf 6}$ & 
${\bf 6}$ & ${\bf 1}$ & ${\bf 1}$ &${\bf 1}$ & ${\bf 6}$ & ${\bf 3}$ & 
${\bf 6}^*$ & ${\bf 6}$ & ${\bf 6}$ \\
U(3)$'$ & 
${\bf 6}^*$ & ${\bf 6}^*$ & ${\bf 1}$ &  ${\bf 6}$ & ${\bf 1}$ & 
${\bf 1}$ & ${\bf 6}$ & ${\bf 6}$ & ${\bf 6}$ & ${\bf 1}$ & ${\bf 3}^*$ &
${\bf 1}$ & ${\bf 1}$ & ${\bf 1}$ \\
$R$ & ${r}_{\Phi\nu}$ & $\bar{r}_{YD}$ & $\bar{r}_{YR}$ & ${r}_{\Phi R}$ & 
${r}_{\Phi u}$ & 
$r_{\Theta e}$ & $r_{\Theta \nu}$ & $\hat{r}_{\Theta\nu}$ &
$r_{\Theta D}$ & $r_{\Theta R}$ & $\hat{r}_{\Theta R}$ &
$\bar{r}_{\Theta u}$ & $\hat{r}_{\Theta u}$ & $r_{\Theta d}$ \\ \hline
\end{tabular}
\begin{tabular}{|cc|cccccc|} \hline
$\Phi_0$ & $X$ & $E$ & $\bar{E}$ & $E'$ & $\bar{E}'$ & 
$E^{\prime\prime}$ & $\bar{E}^{\prime\prime}$ \\ \hline
${\bf 1}$ & ${\bf 1}$ & ${\bf 1}$ & ${\bf 1}$ & ${\bf 1}$ & ${\bf 1}$ & 
${\bf 1}$ & ${\bf 1}$   \\
${\bf 1}$ & ${\bf 1}$ & ${\bf 1}$ & ${\bf 1}$  & ${\bf 1}$ & ${\bf 1}$ &
${\bf 1}$ & ${\bf 1}$ \\
${\bf 3}^*$ & ${\bf 3}$ & ${\bf 6}$ & ${\bf 6}^*$ & ${\bf 3}$ & 
${\bf 3}^*$ & ${\bf 1}$ & ${\bf 1}$ \\
${\bf 3}^*$ & ${\bf 3}$ & ${\bf 1}$ & ${\bf 1}$ & ${\bf 3}^*$ & 
${\bf 3}$ & ${\bf 6}$ & ${\bf 6}^*$ \\
$r_0$ & $\frac{1}{2}(r_E+r^{\prime\prime}_E-1)$ & $r_{E}$ & $1-r_{E}$ &
 $r'_{E}$ & $1-r'_{E}$ & 
$r_{E}^{\prime\prime}$ & $1-r_{E}^{\prime\prime}$   \\ \hline
\end{tabular}
\begin{tabular}{|cccccccccc|} \hline
$E_u$ & $\bar{E}_u$ & $E_d$ & $\bar{E}_d$ & 
$E^{\prime\prime}_u$ & $\bar{E}^{\prime\prime}_u$ & 
$E^{\prime\prime}_d$ & $\bar{E}^{\prime\prime}_d$ & 
$P^d$ & $\bar{P}_d$  \\ \hline
${\bf 1}$ & ${\bf 1}$ & ${\bf 1}$ & ${\bf 1}$ & ${\bf 1}$ & ${\bf 1}$ & 
${\bf 1}$ & ${\bf 1}$ & ${\bf 1}$ & ${\bf 1}$ \\
${\bf 1}$ & ${\bf 1}$ & ${\bf 1}$ & ${\bf 1}$ & ${\bf 1}$ & ${\bf 1}$ & 
${\bf 1}$ & ${\bf 1}$ & ${\bf 1}$ & ${\bf 1}$ \\
${\bf 6}$ & ${\bf 6}^*$ & ${\bf 6}$ & ${\bf 6}^*$ & 
${\bf 1}$ & ${\bf 1}$ & ${\bf 1}$ & ${\bf 1}$ & ${\bf 6}$ & ${\bf 6}^*$  \\
${\bf 1}$ & ${\bf 1}$ & ${\bf 1}$ & ${\bf 1}$ & 
${\bf 6}$ & ${\bf 6}^*$ & ${\bf 6}$ & ${\bf 6}^*$ & 
${\bf 1}$ & ${\bf 1}$ \\
$r_{Eu}$ & $1-r_{Eu}$ & $r_{Ed}$ & $1-r_{Ed}$ & 
$r_{Eu}^{\prime\prime}$ & $1-r_{Eu}^{\prime\prime}$ &
$r_{Ed}^{\prime\prime}$ & $1-r_{Ed}^{\prime\prime}$ 
& $r_{Pd}$ & $1-r_{Pd}$ \\ \hline
\end{tabular}
\end{center}
\end{table}

From Eq.(A.7), we obtain 
$r^{\prime\prime}+ \bar{r}^{\prime\prime}_E =r_E +\bar{r}_E$. 
When we take $R(E^{\prime\prime})+ R(\bar{E}^{\prime\prime}) 
=R(E) +R(\bar{E}) = R(P^d) +R(\bar{P}_d) =1$, 
we can introduce the following superpotential:
$$
W_{E,P} = \frac{\lambda_1}{\Lambda} {\rm Tr}[\bar{E} E \bar{P}_d P_d] 
+\frac{\lambda_2}{\Lambda} {\rm Tr}[\bar{E} E]  {\rm Tr}[\bar{P_d} P_d] ,
\eqno(A.9)
$$
from which we obtain relations  
$\langle E \rangle \langle \bar{E} \rangle \propto {\bf 1}$ and 
$\langle P_d \rangle \langle \bar{P}_d \rangle \propto {\bf 1}$.
We assume following specific solutions of those relations: 
$$
\frac{1}{v_E} \langle E \rangle = \frac{1}{\bar{v}_E} \langle \bar{E} \rangle 
= {\bf 1} , 
\eqno(A.10)
$$
$$
\frac{1}{v_P} \langle P_d \rangle^\dagger = \frac{1}{\bar{v}_P^*} 
\langle \bar{P}_d \rangle = {\rm diag}( e^{-i\phi_1}, e^{-i\phi_2}, 1),
\eqno(A.11)
$$
as the explicit forms of $\langle E \rangle$, $\langle \bar{E} \rangle$ 
and $\langle \bar{P}_d \rangle$.  
We assume similar superpotential forms for $(E, \bar{E})$, 
$(E_u, \bar{E}_u)$, $(E_d, \bar{E_d})$, 
 $(E^{\prime\prime}, \bar{E}^{\prime\prime})$, 
 $(E^{\prime\prime_u}, \bar{E}^{\prime\prime}_u)$,
 $(E^{\prime\prime}_d, \bar{E}^{\prime\prime}_d)$
and $(E', \bar{E}')$. 

The term $\mu_d E_d $ in Eq.(A.6) has been introduced 
in order to adjust the down-quark mass ratio $m_d/m_s$ 
as seen in Sec.3. 
Additional terms like $\mu_d E_d $ in the lepton and up-quark 
sectors do not appear, because we take
$R(E) \neq R(E_d)$ and $R(E_u) \neq (E_d)$.

\vspace{10mm}


%


\begin{thebibliography}{99} 
%
%
\bibitem{CKM} N.~Cabibbo, Phys.~Rev.~Lett. {\bf 10} (1963) 531; 
M.~Kobayashi and T.~Maskawa, Prog.~Theor.~Phys. {\bf 49} 
(1973) 652.
%
\bibitem{PMNS} B.~Pontecorvo, Zh.~Eksp.~Teor.~Fiz. {\bf 33}, 
 549 (1957) and {\bf 34} (1957) 247; 
Z.~Maki, M.~Nakagawa, and S.~Sakata, Prog.~Theor.~Phys. {\bf 28}  
 (1962) 870.
%
\bibitem{O3_PLB09}
Y.~Koide, Phys.~Lett. {\bf B 680} (2009) 76.
%
\bibitem{yukawaon} 
H.~Nishiura and Y.~Koide, Phys.~Rev. {\bf D 83} (2011) 035010.
%
\bibitem{K-N_EPJC12}
Y.~Koide and H.~Nishiura, Euro.~Phys.~J.  {\bf C 72} (2012) 1933.
%
\bibitem{K-N_PLB12}
Y.~Koide and H.~Nishiura, Phys.~Lett. {\bf B 712} (2012) 396.
%
\bibitem{flavon}  C.~D.~Froggatt and H.~B.~Nelsen, Nucl.~Phys. 
{\bf B 147} (1979) 277.
For recent works, for instance, see R.~N.~Mohapatra, 
AIP Conf.~Proc. {\bf 1467} (2012) 7;
A.~J.~Burasu {\it et al}, JHEP {\bf 1203} (2012) 088. 
%
\bibitem{e-mass}
Y.~Sumino, JHEP {\bf 0905} (2009) 075. 
Also see, Y.~Koide, Mod. Phys. Lett. A {\bf 5} (1990) 2319;
E.~Ma, Phys.~Lett. {\bf B 649} (2007) 287.
%
\bibitem{K-N_EPJC13}
Y.~Koide and H.~Nishiura, Eur.~Phys.~J. {\bf C 73} (2013) 2272.
%
%
%
\bibitem{theta13}
K.~Abe {\it et al.} (T2K collaboration),
Phys.~Rev.~Lett. {\bf 107} (2011) 041801;
%
MINOS collaboration, P. Adamson et. al., 
 Phys.~Rev.~Lett. {\bf 107} (2011) 181802;
%
Y. Abe {\it et al.}  (DOUBLE-CHOOZ Collaboration), 
Phys.~Rev.~Lett. {\bf 108}  (2012) 131801;
%
F.~P.~An, {\it et al.} (Daya-Bay collaboration),  
 Phys.~Rev.~Lett. {\bf 108} (2012) 171803;
%
J.~K.~Ahn, {\it et al.} (RENO collaboration),  
 Phys.~Rev.~Lett. {\bf 108} (2012) 191802.
%
%
%
%
\bibitem{PDG2012}
  J.~Beringer {\it et al.},  Particle Data Group,
  Phys.~Rev.\ D {\bf 86} (2012), 0100001.
%
\bibitem{J} 
C.~Jarlskog, Phys.~Rev.~Lett. {\bf 55}, 1839 (1985);
O.~W.~Greenberg,  Phys.~Rev. {\bf D32}, 1841 (1985);
I.~Dunietz, O.~W.~Greenberg and D.-d.~Wu,  Phys.~Rev.~Lett. {\bf 55}, 
2935 (1985);
C.~Hamzaoui and A.~Barroso,  Phys.~Rev. {\bf D33}, 860 (1986).
%
%

%
%
%
\bibitem{MINOS08} P.~Adamson {\it et al.}, MINOS collaboration, 
Phys.~Rev.~Lett. {\bf 101} (2008) 131802. 
%
%
\bibitem{Doi1981} M.~Doi, T.~Kotani, H.~Nishiura, K.~Okuda, and E.~Takasugi,
 Phys.~Lett. {\bf B103} (1981) 219; ibid. {\bf B113} (1982) 513.
%
\bibitem{bb_expmt}
S.~M.~Bilenky and C.~Giunti, 
Mod. Phys. Lett. A {\bf 27} (2012) 1230015.
%
\bibitem{q-mass} Z.-z.~Xing, H.~Zhang, and S.~Zhou, 
{Phys.~Rev.} {\bf D 77} (2008) 113016.
And also see, H.~Fusaoka and Y.~Koide, {Phys. Rev.} 
{\bf D 57} (1998) 3986.
%
\end{thebibliography}
\end{document}